\documentclass[aps,prx,superscriptaddress,twocolumn,eprint,amsmath,amssymb,longbibliography,nofootinbib]{revtex4-1}

\usepackage{amsmath,amsthm,amsfonts,amssymb,amscd}
\usepackage{fullpage}
\usepackage{lastpage}
\usepackage{graphicx}
\usepackage{hyperref}
\usepackage{wrapfig}
\usepackage{enumerate}
\usepackage{fancyhdr}
\usepackage{mathrsfs}
\usepackage{mathtools}
\usepackage{braket}
\usepackage{slashed}
\usepackage{bm}
\usepackage[dvipsnames]{xcolor}
\usepackage{float}
\usepackage[margin=0.55in]{geometry}
\usepackage[english]{babel}
\usepackage{xcolor,cancel}
 
\usepackage{subcaption}
\usepackage{comment}
\usepackage{tikz}
\usetikzlibrary{calc}
\usetikzlibrary{decorations.markings}
\usepackage[section]{placeins} % should prevent floats from being moved to different section

\newcommand{\be}{\begin{equation}}
\newcommand{\ee}{\end{equation}}
\newcommand{\bea}{\begin{eqnarray}}
\newcommand{\eea}{\end{eqnarray}}

\newcommand{\br}{\mathbf{r}}

\def\onedot{$\mathsurround0pt\ldotp$}
\def\cddot{% two dots stacked vertically
  \mathbin{\vcenter{\baselineskip.67ex
    \hbox{\onedot}\hbox{\onedot}}%
  }}%

\begin{document}

\title{Poisson-bracket formulation of the dynamics of fluids of deformable particles}
\author{Arthur Hernandez}
\email[Corresponding author:\ ]{arthurhernandez@umail.ucsb.edu}
\affiliation{Department of Physics, University of California Santa Barbara, Santa Barbara, CA 93106, USA}
\author{M.~Cristina Marchetti}
\affiliation{Department of Physics, University of California Santa Barbara, Santa Barbara, CA 93106, USA}
\email[\ ]{cmarchetti@ucsb.edu}
%\date{}                                           % Activate to display a given date or no date

\begin{abstract}
    {Using the Poisson bracket method, we derive continuum equations for a fluid of deformable particles in two dimensions. Particle shape is quantified in terms of two continuum fields: an anisotropy density field that captures the deformations of individual particles from regular shapes and a shape tensor density field that quantifies both particle elongation and nematic alignment of elongated shapes. We explicitly consider the example of  a dense biological tissue as described by the Vertex model energy, where cell shape has been proposed as a structural order parameter for a liquid-solid transition. The hydrodynamic model of biological tissue proposed here captures the coupling of cell shape to flow, and provides a starting point for modeling the rheology of dense tissue. }
\end{abstract}

\maketitle
\section{Introduction}
%\subsection{}

Many extended systems, such as biological tissue~\cite{lecuit2007cell}, foams~\cite{durian1995foam,cohen2013flow}, emulsions~\cite{mattsson2009soft,vlassopoulos2014tunable}, and colloidal suspensions~\cite{mattsson2009soft}
 can be described as collections of deformable particles. A variety of mesoscopic models have been developed to examine the role of particle shape  on the structure and rheology of these soft materials. 
 
 %Particle-based models provide a natural framework for incorporating cell motility and quantifying the role of variation in the density of cells in controlling their collective behavior~\cite{szabo2006phase,henkes2011active,hakim2017collective}. 
 Cellular Potts models~\cite{graner1992simulation,kabla2012collective} and Vertex and Voronoi models~\cite{honda2004three,hufnagel2007mechanism,farhadifar2007influence} have been successfully used to describe dry foams and confluent layers of biological tissue, where cells completely cover the plane with no gaps,  with  extensions to three dimensions~\cite{hannezo2014theory,murisic2015discrete}. These models describe cells in confluent tissues  as tightly packed irregular polygons covering the plane and predict   a jamming-unjamming transition  tuned by a target cell shape that captures the interplay of cortex contractility and cell-cell adhesion, with the mean cell shape  serving as a metric for tissue fluidity~\cite{bi2014energy,bi2015density,bi2016motility}. Vertex and Voronoi models do not, however, have a natural extension to situations where the cell packing fraction is below one, although gaps between cells have been incorporated in recent work~\cite{teomy2018confluent,kim2020embryonic}. In contrast, both  particle deformability and density variations can be incorporated in  multi-phase field models and in models of deformable particles~\cite{boromand2019role},   which  have been used to examine solid-liquid transitions as a functions of both particle shape and density. 
 
Less well developed are continuum descriptions of the rheology of materials where the constituents can change their shape. An important example is the classic work by Doi and Ohta that describes the dynamics of the interface between two immiscible fluids under shear, incorporating formation, rupture and deformation of droplets~\cite{doi1991dynamics}.   Continuum mechanics of confluent tissue have  been constructed phenomenologically and employed to connect structure and mechanics in \emph{Drosophila} development~\cite{sagner2012establishment,popovic2017}. Ishihara and collaborators formulated a continuum model that couples cell shape to mechanical deformations at the tissue scale~\cite{ishihara2017cells}. Their work, however, only captures \emph{simultaneous} cell anisotropy and alignment of elongated cell shapes, without distinguishing between a tissue  where cell shapes are on average isotropic and one where cells are on average anisotropic, but not aligned, as observed in simulations of Vertex/Voronoi models~\cite{bi2016motility,yang2017correlating}.   It is in fact the single-cell anisotropy that provides an order parameter for cell jamming in Vertex and Voronoi models~\cite{bi2014energy,bi2015density,bi2016motility}, where fluid states of elongated cells are obtained without nematic order of elongated cells. The importance of this distinction in a continuum theory of tissue mechanics was highlighted recently in work by one of us and collaborators~\cite{czajkowski2018hydrodynamics}.

In this paper we adopt the Poisson-bracket formulation~\cite{forster1974microscopic} to obtain continuum equations for a fluid of deformable particles in two dimensions. This method has the advantage of providing a systematic derivation of the reversible part of the hydrodynamic equations once the continuum fields have been identified. Our approach is inspired by work by Stark and Lubensky~\cite{Stark2003,Stark_2005} who used the Poisson-bracket approach to derive the hydrodynamics of a nematic liquid crystal. As in liquid crystals, we identify both a continuum scalar field that quantifies fluctuations of individual cell shape and a cell shape tensor field that captures both cell elongation and alignment. An important difference is that, while in passive liquid crystals molecular shape fluctuations decay on fast (non-hydrodynamic) time scales, in a tissue cell shape is the order parameter for the rigidity transitions, hence cell-shape fluctuations are long-lived near the transition and must be incorporated in a hydrodynamic model.  The equations derived here provide a  continuum model for collections of interacting deformable ``particles'' and can be adapted to describe both confluent and non-confluent systems. 

The paper is organized as follows.
In Section \ref{sec:fields} we provide the microscopic definition of the continuum fields used in the hydrodynamic model. In Section \ref{sec:PB} we briefly summarize the Poisson Bracket (PB) method and the calculation of the various PBs (with details given in Appendix~\ref{app:PB}), and discuss the reactive and dissipative contributions to the coarse-grained dynamics. The final continuum equations are displayed in Section \ref{sec:final}. In Section \ref{sec:tissue} we discuss the form of the continuum equations for the specific case of a cellular tissue, and conclude with a brief discussion in Section \ref{sec:end}. Details of the derivation of the PBs and of the mean-field free-energy of the Vertex model are given in Appendices.

\section{Continuum fields}
\label{sec:fields}
\begin{figure}[h!]
\centering
\includegraphics[width=6cm]{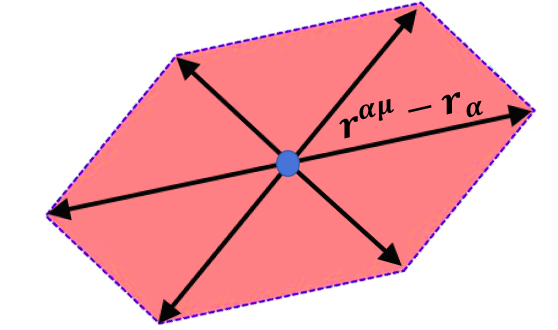}
\caption{A deformable particle (referred to as a `cell') is described as an $n$-sided irregular polygon defined by the positions $\mathbf{r}_{\alpha\mu}$ of its vertices, for $\mu=1,\cdots,n$, relative to the location of the centroid $\mathbf{r}_\alpha$ of the polygon.}
\label{fig:shape}
\end{figure}

We consider a fluid whose constituents are $N$ extended particles of arbitrary shape. The contour of each particle, referred to below as a `cell', is described by a polygonal shape joining $n$ vertices located at $\br_\mu^\alpha$, where $\mu=1,2,\cdots, n$ labels the vertices and $\alpha=1,2,\cdots,N$ labels the cells,  as shown in Fig.~\ref{fig:shape}. Each cell has  a total mass $m_c$, which we assume equally distributed among the $n$ vertices. 
The shape of each cell is described by a shape tensor defined as

\be
{G}^\alpha_{ij}=\frac{1}{n}\sum_{\mu=1}^{n}\Delta x_i^{\alpha\mu}\Delta x_j^{\alpha\mu}\;,
\label{eq:G-alpha}
\ee
where $\Delta\br^{\alpha\mu}=\br^{\alpha\mu}-\br_\alpha$,  with $\br_\alpha=\frac{1}{n}\sum_\mu\br^{\alpha\mu}$, and Latin indices $i,j$ denote components. 
We define microscopic mass, momentum and cell shape density fields as
\begin{align}
&\hat{\rho}(\br,t)=\sum_{\alpha\mu}m~\delta(\br-\br^{\alpha\mu}(t))\;, \label{eq:rho-hat}\\
&\hat{\mathbf{g}}(\br,t)=\sum_{\alpha\mu}m\dot{\br}^{\alpha\mu}~\delta(\br-\br^{\alpha\mu}(t))\;, \label{eq:g-hat}\\
&\hat{G}_{ij}(\br,t)=\sum_\alpha{G}^\alpha_{ij}~\delta(\br-\br_\alpha(t))\;. \label{eq:G-hat}
\end{align}
with $m=m_c/n$. Coarse grained quantities are then defined as $\rho(\br,t)=\left[\hat{\rho}(\br,t)\right]_c$, $\mathbf{g}(\br,t)=\left[\hat{\mathbf{g}}(\br,t)\right]_c$ and $G_{ij}(\br,t)=\left[\hat{G}_{ij}(\br,t)\right]_c$ and correspond to macroscopic continuum  fields describing the system on length scales large compared to both the size of the particles and their mean separation. Note that since the microscopic cell-shape tensor $\hat{\mathbf{G}}^\alpha$ has dimensions of length squared, the density of cellular shape tensor $G_{ij}$ is dimensionless.
As we will see below, the trace of the shape tensor density provides a measure of the density of cell perimeter, while its traceless part, $\tilde{G}_{ij}=G_{ij}-\frac12\delta_{ij}\text{Tr}[\mathbf{G}]$, captures both cell anisotropy and local alignment of elongated cells.

The cellular shape tensor can be written in terms of its eigenvalues as
\be
{G}^\alpha_{ij}=\frac12\left(\lambda_1^\alpha+\lambda_2^\alpha\right)\delta_{ij}+\left(\lambda_1^\alpha-\lambda_2^\alpha\right)
\left(\hat{\nu}_{i}^\alpha\hat{\nu}_j^\alpha-\frac12\delta_{ij}\right)\;,
\label{eq:Ga_eig}
\ee
where $\lambda_1^\alpha>\lambda_2^\alpha$ and $\hat{\bm\nu}^\alpha$ is the eigenvector of the largest eigenvalue.
For regular  $n-$sided polygons, the shape tensor is diagonal with $\lambda_1^\alpha= \lambda_2^\alpha$. In this case the cell area $A_\alpha^{(n)}$ and perimeter $P_\alpha^{(n)}$   can be expressed in terms of the invariants of the tensor $\mathbf{G}^\alpha$ as
\begin{align}
&A_\alpha^{(n)}=n\sin\left(\frac{2\pi}{n}\right)\sqrt{\det[\mathbf{G}^\alpha]}\;, \label{eq:An}\\
&P_\alpha^{(n)}=2n\sin\left(\frac{\pi}{n}\right)\sqrt{\text{Tr}[\mathbf{G}^\alpha]}\;, \label{eq:Pn}
\end{align}

Single cell anisotropy is measured by $\Delta_\alpha=\lambda_1^\alpha-\lambda_2^\alpha$
which vanishes for regular polygons.
To quantify single-cell elongation independently of alignment of elongated cells, we follow \cite{czajkowski2018hydrodynamics}, albeit with a slightly different definition of the shape tensor, and introduce an anisotropy density field defined as
\be
\hat{M}(\br,t)=\sum_\alpha \Delta_\alpha\delta(\br-\br_\alpha(t))
\label{eq:m-hat}
\ee
and the associated coarse grained field $M(\br,t)=[\hat{M}(\br,t)]_c$.
Work on Vertex/Voronoi models of confluent biological tissue, as well as multiphase fields models, has demonstrated the correlation between tissue fluidity and anisotropy of single cell shape, as quantified here by $M$. In Vertex models, this anisotropy  provides an order parameter for the solid-liquid transition~\cite{bi2015density,bi2016motility}.

In the following, we construct hydrodynamic equations for a fluid of deformable particles that couple structural changes encoded in cell shape and alignment of elongated cells to flow. The dynamics of the fluid on scales large compared to the cell size and mean cell separation is described in terms of a few continuum fields: the mass density $\rho$, the momentum density $\mathbf{g}$, the single-cell anisotropy density $M$ and the cell-shape tensor density $G_{ij}$.

\section{Poisson-Bracket formulation of continuum dynamics}
\label{sec:PB}
Here we briefly summarize the Poisson-Bracket (PB) formalism. Consider a system whose microscopic dynamics is determined by canonically conjugate positions $\mathbf{r}^\alpha$ and momenta $\mathbf{p}^\alpha$. We describe the dynamics in terms of a few microscopic density fields $\hat\Psi^a(\br,t;\{\mathbf{r}^\alpha\},\{\mathbf{p}^\alpha\})$, for $a=1,2,\cdots$.  These fields are chosen to be be either hydrodynamic fields associated with conserved quantities, broken symmetry fields, or quasi-hydrodynamic fields that decay on times scales large compared to microscopic ones. In the specific case of interest here $\{\hat\Psi^a\}=(\hat\rho, \mathbf{\hat{g}},\hat{G}_{ij},\hat{M})$. The dynamics of the corresponding coarse-grained fields $\Psi^a(\br,t)=[\hat\Psi^a(\br,t;\{\mathbf{r}^\alpha\},\{\mathbf{p}^\alpha\})]_c$ is governed by the equations
\be
\partial_t\Psi^a(\br,t)=V^a(\br,t)+D^a(\br,t)\;,
\label{eq:dyn}
\ee
where $V^a$ and $D^a$ represent the non-dissipative and dissipative parts of the dynamics, respectively.
The reactive term $V^a$ is given by
\be
V^a(\br)=-\int_{\br'}\{\Psi^a(\br),\Psi^b(\br')\}\frac{\delta\mathcal{F}}{\delta\Psi^b(\br')}\;,
\label{eq:Va}
\ee
where $\mathcal{F}[\{\Psi^a\}]$ is a phenomenological free energy,
\be
\{\Psi^a(\br),\Psi^b(\br')\}=\left[\{\hat\Psi^a(\br),\hat\Psi^b(\br')\}\right]_c\;,
\label{eq:PB-cg}
\ee
and
\bea
\{\hat\Psi^a(\br),\hat\Psi^b(\br')\}
=&&\sum_{\alpha i}\Big( \frac{\partial\hat\Psi^a(\br)}{\partial p_i^\alpha}\frac{\partial\hat\Psi^b(\br')}{\partial r_i^\alpha}\nonumber\\
&&-\frac{\partial\hat\Psi^a(\br)}{\partial r_i^\alpha}\frac{\partial\hat\Psi^b(\br')}{\partial p_i^\alpha}\Big)\;.
\label{eq:PB}
\eea
Finally, the dissipative term in the kinetic equation is controlled by all the neglected microscopic degrees of freedom and can be written as
\be
D^a(\br)=-\Gamma^{ab}\frac{\delta\mathcal{F}}{\delta\Psi^b(\br)}\;.
\label{eq:Da}
\ee
The dissipation tensor $\Gamma^{ab}$ is in general a functional of the $\{\Psi^a\}$ and their gradients. It is a phenomenological quantity controlled by the requirement
that $\partial_t\Psi^a$ can only couple to driving forces $\frac{\delta\mathcal{F}}{\delta\Psi^b(\br)}$ that have different sign under time reversal, to guarantee that such terms describe dissipation. In equilibrium it is a symmetric tensor and must obey Onsager's principle~\cite{de1951thermodynamics}.

\subsection{Poisson brackets}
The calculation of the PB of mass and momentum density is straightforward and can be found in the literature~\cite{Stark2003}, with the result 
\begin{align}
&\{\rho(\br),g_i(\br')\}=\rho(\br')\partial_i\delta(\br-\br')\;,\nonumber\\
&\{g_i(\br),g_j(\br')\}= -\partial_i' [\delta(\br-\br') g_j(\br^{\prime})]+\partial_j \delta (\br-\br')g_i(\br')\;,
\end{align}

The main PBs to be calculated here are those involving the fields describing cellular shape. The details of the derivation are shown in Appendix \ref{app:PB}, 
 with the result
\bea
\{G_{ij}(\br),g_k(\br')\}&&=\partial_k[G_{ij}(\br')\delta(\br-\br')]\notag\\
&&-\left[G_{il}(\br)\delta_{jk}+G_{jl}(\br)\delta_{ik}\right]\partial_l\delta (\br-\br')\;,
\label{eq:PB-Gij}
\eea
\be
\{M(\br),g_i(\br')\}=[\partial_i M(\br)]\delta(\br-\br')-\frac{2R(\br)}{M(\br)}\tilde{G}_{ij}(\br)\partial_j\delta(\br-\br')\;.
\ee
To calculate  $\{M(\br),g_i(\br')\}$ we have used the identity    
$\tilde{G}_{ik}^\alpha \tilde{G}_{kj}^{\alpha} = \frac{\Delta_{\alpha}^2}{4}\delta_{ij}$, 
where the tilde denotes the traceless part of
ny rank-2 tensor, 
$\tilde{G}_{ij} = G_{ij}-\frac12\delta_{ij} \text{Tr}[\mathbf{G}]$.\\
This allows us to write
\be
\Delta_\alpha\{\Delta_\alpha\delta(\br-\br^\alpha),g_i(\br')\}=2\tilde{G}_{kl}^\alpha\{\tilde{G}_{kl}^\alpha\delta(\br-\br^\alpha),g_i(\br')\}\;.
\ee

Finally, the other PBs can be obtained using the identity
\be
\{\Psi_n(\br),\Psi_m(\br')\}=-\{\Psi_m(\br'),\Psi_n(\br)\}\;.
\ee

\subsection{Reactive terms}
To evaluate the various contributions to the continuum dynamics, we need to specify the free energy of the system. In general, this has the form
\bea
\mathcal{F}&=&\mathcal{F}_K+\mathcal{F}_V\notag\\
&=&\int_\br \left[\frac{\mathbf{g}^2}{2\rho}+f(\rho,G_{ij},M)\right]\;,
\label{eq:F}
\eea
where the first term is the kinetic part and  the free energy density $f$ depends on the fields and their gradients.

Using the expressions for the Poisson brackets we can then evaluate the reactive terms $V^a$, with the result
\bea
V^\rho&=&-\bm\nabla\cdot(\rho\mathbf{v})\;,
\label{eq:Vrho}\\
V^g_i&=&-\partial_j(\rho v_iv_j)-\rho\partial_i\frac{\delta\mathcal{F}_V}{\delta\rho}
+(\partial_iM)\frac{\delta\mathcal{F}_V}{\delta M}+\left(\partial_iG_{kl}\right)\frac{\delta\mathcal{F}_V}{\delta G_{kl}}\notag\\
&& +\partial_j\left(2G_{jk}\frac{\delta\mathcal{F}_V}{\delta G_{ik}}-\delta_{ij}G_{kl}\frac{\delta\mathcal{F}_V}{\delta G_{kl}}\right)\notag\\
%+(\partial_i M) \left(\frac{\delta\mathcal{F}_V}{\delta M}}\right) 
&&+ 2\partial_j \left( \frac{R}{M}\tilde{G}_{ij} \frac{\delta \mathcal{F}_V}{\delta M}\right) \;,
\label{eq:Vg}\\
V^G_{ij}&=&-\bm\nabla\cdot (G_{ij}\mathbf{v})+G_{ik}\partial_kv_j+G_{jk}\partial_kv_i\ \;,
\label{eq:VGij}\\
V^M&=&  -\mathbf{v}\cdot\bm\nabla M+\frac{2 R}{M}\tilde{G}_{ij}\partial_iv_j\;,
\label{eq:VM}
\eea
where we have defined
\be
R(\br,t)=Tr[\mathbf{G}(\br,t)]\;.
\ee
The field $R$ is essentially a measure of cell perimeter density.

The elastic and density couplings in Eq.~\eqref{eq:Vg} can be rewritten  in a more familiar form as gradients of pressure and of an elastic stress. The details can be found in Appendix \ref{app:pressure}, where it is shown that we can write
\be
-\rho\partial_i\frac{\delta\mathcal{F}_V}{\delta\rho}
+(\partial_iM)\frac{\delta\mathcal{F}_V}{\delta M}+(\partial_iG_{kl})\frac{\delta\mathcal{F}_V}{\delta G_{kl}}=-\partial_ip+\partial_j\sigma_{ij}^E
\ee
where  the pressure $p$  and  the elastic stress $\sigma_{ij}^E$, that plays the role of the Erickssen stress of nematic liquid crystals, are given by
\bea
p&=&\rho\frac{\partial f}{\partial\rho}-f
\label{eq:pressure}\;,\\
\sigma_{ij}^E&=&
%-\frac{\partial f}{\partial \nabla_i \rho} \nabla_j \rho 
- \frac{\partial f}{\partial \nabla_j M} \nabla_i M
- \frac{\partial f}{\partial \nabla_j G_{kl}}\nabla_i G_{kl}\;.
\label{eq:sigmaE}
\eea
The last two terms in Eq.\eqref{eq:Vg} correspond to gradients of a reactive elastic stress $\sigma^G_{ij}$, given by
\bea
\sigma^G_{ij}= 
2\frac{R}{M}\tilde{G}_{ij} \frac{\delta \mathcal{F}_V}{\delta M} +   2G_{jk}\frac{\delta \mathcal{F}_V}{\delta G_{ik}}-\delta_{ij}
G_{kl}\frac{\delta \mathcal{F}_V}{\delta G_{kl}}\;.
%-\left[\frac{\partial f}{\partial \nabla_i \rho} \nabla_j \rho + \frac{\partial f}{\partial \nabla_i M}\nabla_j M + \frac{\partial f}{\partial \nabla_i G_{kl}}\nabla_j G_{kl}\right]
%- \frac{\partial f}{\partial \nabla_i M} \nabla_j M
%- ( G_{mn}) \frac{\partial f}{\partial \nabla_j G_{mn}}
\eea
 The  reactive term for the momentum density equation can then be written as
\be
V^g_i= -\partial_j (\rho v_i v_j) - \partial_i p + \partial_{j} \left(\sigma^G_{ij} +\sigma^E_{ij} \right)\;.
\ee

%Note that $G_{ij}$ here plays the role of the conformation tensor in the theory of viscoelastic polymer suspensions~ \cite{beris1994thermodynamics}.

\subsection{Dissipative terms}
There is no dissipative term for the mass density $\rho$ if it is conserved. 

Dissipative terms in the momentum equation must be odd under time reversal and hence must couple to gradients of velocity. In general, shape anisotropy and alignment of elongated cells will entail anisotropic viscosity coefficients, as in liquid crystals. For simplicity, here we only introduce two viscosities to account for shear ($\eta$) and bulk ($\eta_b$) deformations and write
\be
D^g_i=\partial_j\sigma_{ij}^D\;,
\label{eq:Dg}
\ee
with
\be
\sigma_{ij}^D=2\eta D_{ij}+\eta_b\delta_{ij}\bm\nabla\cdot\mathbf{v}\;,
\label{eq:sigmav}
\ee
where $D_{ij}$ is the symmetrized and traceless rate of strain tensor,
\be
D_{ij}=\frac12(\partial_iv_j+\partial_jv_i-\delta_{ij}\bm\nabla\cdot\mathbf{v})\;.
\label{eq:Dij}
\ee

Dissipative couplings in the equations for the shape density tensor $G_{ij}$ and the shape anisotropy field $M$ must be even under time reversal and hence can couple to  $M$, $G_{ij}$ and their gradients. Dissipation will arise from topological rearrangements, as well as from birth/death events when density conservation is broken. In general we can write
\bea
D^G_{ij}&=&-\Gamma^{GG}_{ijkl}\frac{\delta\mathcal{F}_V}{\delta G_{kl}}-\Gamma^{GM}_{ij}\frac{\delta\mathcal{F}_V}{\delta M}\;,\\
D^M&=&-\Gamma^{MM}\frac{\delta\mathcal{F}_V}{\delta M}-\Gamma^{MG}_{ij}\frac{\delta\mathcal{F}_V}{\delta G_{ij}}\;.
\eea
The kinetic coefficients $\Gamma^{ab}$ can generally depend on the shape tensor and anisotropy density field. To linear order in these fields, a general form is given by
\bea
\Gamma^{GG}_{ijkl}&=&\frac{M}{2\gamma_{G}}\left(\delta_{ik}\delta_{jl}+\delta_{jk}\delta_{il}\right)\notag\\
&&+\frac{1}{\gamma_1}\left(\delta_{ik}G_{jl}
+\delta_{jk}G_{il}+\delta_{il}G_{jk}+\delta_{jl}G_{ik}\right)\;,
\label{eq:GammaGG}\\
\Gamma^{GM}_{ij}&=&\Gamma^{GM}_{ij}=\frac{G_{ij}}{\gamma_2}
\;,
\label{eq:GammaGM}\\
\Gamma^{MM}&=&\frac{1}{\gamma_M}+\frac{M}{\gamma_3}
\;,\label{eq:GammaMM}
\eea
where the kinetic coefficients  $\gamma_i$, for $i=G,M,1,2,3$, encode the characteristic time scales of dissipative processes.
For simplicity we have assumed $\Gamma^{GM}_{ij}=\Gamma^{GM}_{ij}$ although in general the parameters controlling the relaxation in these terms could differ. 
Note that the second term in Eq.~\eqref{eq:GammaGG} has the form introduced in Ref.~\cite{milner1993dynamical} for the kinetic coefficient describing the relaxation of the conformation tensor in a polymer suspension.

\section{Final equations}
\label{sec:final}
Putting it all together, we now write  the final form of the equations we have obtained. It is convenient to write 
\be
\partial_iv_j=D_{ij}+\omega_{ij}+\frac12\delta_{ij}\bm\nabla\cdot\mathbf{v}
\ee
where $D_{ij}$ is the  rate of strain tensor given in Eq.~\eqref{eq:Dij} and $\omega_{ij}$ is the vorticity,
\bea
\omega_{ij}=\frac12(\partial_iv_j-\partial_jv_i)\;.
\eea
The set of continuum equations for our fluid of deformable cells is then given by
\bea
&&\partial_t\rho =-\bm\nabla\cdot\rho\mathbf{v},
\label{eq:rhoeq}\\
&&\rho\left(\partial_t+\mathbf{v}\cdot\bm\nabla\right)v_i= - \partial_i p + \partial_j\left(\sigma^G_{ij}+ \sigma^E_{ij} + \sigma^{D}_{ij}\right),
\label{eq:geq}\\
&&\frac{d}{dt}M= \frac{2R}{M}\tilde{G}_{ij} D_{ij} -\Gamma^{MM} \frac{\delta \mathcal{F}_V}{\delta M}
-\Gamma^{MG}_{ij} \frac{\delta \mathcal{F}_V}{\delta G_{ij}},\label{eq:Meq}\\
&&\frac{D}{Dt}G_{ij}=G_{ik}D_{kj}+D_{ik}G_{kj}-\Gamma^{GG}_{ijkl}\frac{\delta \mathcal{F}_V}{\delta G_{kl}}-\Gamma^{GM}_{ij}\frac{\delta \mathcal{F}_V}{\delta M},
\label{eq:Geq}
\eea
where 
\begin{align}
&\frac{d}{dt}=\partial_t+\mathbf{v}\cdot\bm\nabla\;,\notag\\
&\frac{D}{Dt}= \frac{d}{dt}+[\bm\omega,\large\cdot]
\end{align}
are the convective derivative, and the convective and corotational derivative.

The equation for the shape tensor $G_{ij}$ contains couplings to flow vorticity and strain rate which control the tendency of extended and deformable particles to rotate with flow and align with streamlines. 
The shape tensor $G_{ij}$ plays a role similar to that of the conformation tensor in a polymer suspension~\cite{beris1994thermodynamics}. In fact, if we ignore the additional anisotropy density field $M$, the equations derived here for a fluid of deformable particles have the same structure as a one-fluid model of viscoelastic polymer solutions~\cite{milner1993dynamical}.
Unlike in models of polymer suspensions, however, the coefficient of the coupling to strain rate, known in that context as the slip parameter~\cite{beris1994thermodynamics}, is found  to be simply equal to $1$ in our PB formulation.

It is also convenient to separate the dynamics of $G_{ij}$ in that of its trace and deviatoric part.
The corresponding equations are given by
\bea
\frac{d}{dt}R&=&2\tilde{G}_{ij}D_{ji}-\Gamma^{GG}_{iikl}\frac{\delta \mathcal{F}_V}{\delta G_{kl}}-\Gamma^{GM}_{ii}\frac{\delta \mathcal{F}_V}{\delta M},\label{eq:R}\\
\frac{D}{Dt}\tilde{G}_{ij}&=&R D_{ij}+\tilde{G}_{ik}D_{kj}+D_{ik}\tilde{G}_{kj}-\delta_{ij}\tilde{G}_{kl}D_{kl}\notag\\
&&-\left[\bm\Gamma^{GG}\cddot\frac{\delta \mathcal{F}_V}{\delta \mathbf{G}}\right]_{ij}^{ST}-\left[\bm\Gamma^{GM}\right]_{ij}^{ST}\frac{\delta \mathcal{F}_V}{\delta M}\;.
\eea
where $[...]^{ST}$ denotes the symmetrized and traceless part of the tensor.
These equations need to be completed by an expression for the free energy $\mathcal{F}_V$ in terms of the shape tensor and anisotropy density field. Such an expression of course depends on the system of interest. In the next section we consider the specific case of a model of dense biological tissues.

\section{Cellular tissue}
\label{sec:tissue}

Confluent biological tissue, where cells are tightly packed, with no intervening gaps, have been modeled extensively using Vertex or Voronoi models that describe cells as irregular polygons tesselating the plane~\cite{honda2004three,hufnagel2007mechanism,farhadifar2007influence}.
%(see Fig.~\ref{fig:VM}). 
The behavior of the tissue is controlled by an energy that describes the tendency of each cell to adjust   its area $A_a$ and perimeter $P_a$ to target values $A_0$ and $P_0$, given by
\be
E_V=\sum_\alpha\left[\frac{\kappa_A}{2}\left(A_\alpha-A_0\right)^2+\frac{\kappa_P}{2}\left(P_\alpha-P_0\right)^2\right]\;,
\ee
with $\kappa_A$ and $\kappa_P$ stiffness parameters. 
%\begin{figure}[h!]
%\centering
%\includegraphics[width=6cm]{vertexmodel.png}
%\caption{Vertex Model of cells as a polygonal tessellation of the plane. }
%\label{fig:VM}
%\end{figure}
The first term arises from tissue incompressibility in three dimensions and the second captures the interplay of cell-cell adhesion and cortical contractility. 
By scaling lengths with $\sqrt{A_0}$ and energies with $\kappa_A A_0^2$, the scaled energy of each cell is given by
\be
\epsilon_a=\frac12\left(a_a-1\right)^2+\frac{r}{2}\left(p_a-p_0\right)^2\;,
\ee
with $p_0=P_0/\sqrt{A_0}$ the target shape parameter and $r=\kappa_P/(\kappa_A A_0)$.

Numerical studies of this energy have identified a rigidity transition at a critical value $p_0^*$ of the target shape parameter between a rigid, solid-like state for $p_0<p_0^*$ to a fluid state for $p_0>p_0^*$. Single-cell anisotropy as quantified by the mean cell-shape index $q=\langle P_\alpha/\sqrt{A_\alpha}\rangle$, with the brackets denoting an average over all cells, provides an order parameter for the transition. Czajkowski \emph{et al.}~\cite{czajkowski2018hydrodynamics} derived a mean-field model of this rigidity transition, albeit using a different definition of the cell shape tensor $G_{ij}^\alpha$ as compared to the one used here. The derivation carried out with our definition is outlined in Appendix ~\ref{app:MFT}. The result is a quartic Landau-type free energy density $f_M$ where the the cell shape anisotropy density  $M$ plays the role of an order parameter, given by
%
%Since the derivation of Ref.~uses a slightly different definition of the shape tensor.  \mcm{we need to redo this using our definitions.} After rewriting the single-cell energy $\epsilon_a$ in terms of a cell anisotropy parameter 
%$m_a=(\sqrt{\lambda_1^a}-\sqrt{\lambda_2^a})/(\sqrt{\lambda_1^a}+\sqrt{\lambda_2^a})$ and the perimeter $P_\alpha=\sqrt{\lambda_1^a}+\sqrt{\lambda_2^a}$,
%they showed that $\epsilon_a$ can be approximated by a Landau-type free energy density $f_m$ that reproduces the liquid-solid transition given by
\be
f_M=\frac{\alpha(p_0)}{2}M^2+\frac{\beta}{4}M^4\;,
\label{eq:fm}
\ee
where $\alpha(p_0)$ vanishes at $p_0=p_0^*$ and $\beta>0$. The definition of the shape tensor of individual cells only affects the precise values of these parameters that also  depend on the reference polygonal shape, but does not change the form of the free energy density nor the value of $p_0^*$.
The free energy given in Eq.~\eqref{eq:fm} is obtained by assuming  small deformations from regular polygons and constant
 cell perimeter. It predicts a mean-field transition at $\alpha=0$ from a state where cells are isotropic ($M=0$) $\alpha<0$ or $p_0<p_0^*$ (the solid state) 
 to a state where cells are anisotropic ($M=\sqrt{-\alpha/\beta}$)  for $\alpha<0$ or $p_0>p_0^*$ (the liquid state).
 
 This work suggests a phenomenological free energy for a confluent tissue that  captures both fluctuations in the cell anisotropy density  $M$ that quantifies the liquid-solid transition  and the shape tensor density $\tilde{G}_{ij}$ that quantifies   alignment of elongated cell  as
 \bea
 \mathcal{F}_c=\int_\br\bigg[&&\frac{\alpha(p_0)}{2}M^2+\frac{\beta}{4}M^4+\frac{K}{2}(\bm\nabla M)^2\notag\\
 &&+\frac{\chi}{2}Tr[\tilde{\mathbf{G}}^2]+
 \frac{K_G}{2}(\partial_j\tilde{G}_{ik})^2\bigg]\;.
 \eea
We do not include terms of order $Tr[\tilde{\mathbf{G}}^2]^2$ as we do not expect any nematic order of cellular shapes in the absence of externally applied or actively generated internal stresses.  Also, we have assumed constant cell perimeter, corresponding to $R=Tr[\mathbf{G}]=$constant. In general, the various parameters in $\mathcal{F}_c$ will depend on $R$.

It is important to stress that $\tilde{G}_{ij}$ and $M$ are not independent. The traceless tensor $\tilde{G}_{ij}$ can  be written as
\be
\tilde{G}_{ij}=S_G\left(n_in_j-\frac12\delta_{ij}\right)\;,
\ee
which defines the director field $\mathbf{n}(\mathbf{r},t)$ associated with alignment of elongated cells and the magnitude $S$ of orientational order. Cell alignment can only occur if cells are elongated ($M\not=0$), hence $S_G(M)$ must vanish when $M=0$. We assume $S_G=MS$, where $S$ plays the role of a nematic order parameter for orientational order of elongated cells. Clearly, $S$ is defined only in states where $M$ is finite.

Cell sheets commonly interact with a frictional substrate that eliminates momentum conservation. Frictional drag with the substrate generally exceeds inertial forces, and the Navier-Stokes equation for the momentum is replaced by a Stokes equation quantifying force balance on each fluid element. Within this overdamped limit, and considering a minimal form for the various dissipative kinetic coefficients, the tissue dynamics is governed by the following equations
\be
\partial_t\rho =-\bm\nabla\cdot(\rho\mathbf{v}),
\label{eq:rhoeq-tissue}
\ee
\be
\xi v_i= - \partial_i p + \partial_j\left(\sigma^G_{ij}+ \sigma^E_{ij} + \sigma^{D}_{ij}\right),
\label{eq:veq}
\ee
\be
\frac{d}{dt}M= 2\frac{R}{M}\tilde{G}_{ij} D_{ij} -\frac{1}{\gamma_M}\frac{\delta\mathcal{F}_c}{\delta M}-\frac{\tilde{G}_{ij}}{\gamma_2}\frac{\delta\mathcal{F}_c}{\delta\tilde{G}_{ij}}\;,
\label{eq:Meq-tissue}
\ee
\bea
\frac{D}{Dt}\tilde{G}_{ij} &=&R D_{ij}+\tilde{G}_{ik}D_{kj}+D_{ik}\tilde{G}_{kj}-\delta_{ij}
\tilde{G}_{kl} D_{kl} \notag\\
&-&\frac{M}{\gamma_G}\frac{\delta \mathcal{F}_{c} }{\delta \tilde{G}_{ij}}-\frac{\tilde{G}_{ij}}{\gamma_2} \frac{\delta \mathcal{F}_c}{\delta M} \;,
\label{eq:Gij-tissue}
\eea
where $\xi$ is the frictional drag and  
\bea
\frac{\delta\mathcal{F}_c}{\delta M}&=&
\left[\alpha+\beta M^2\right]M-K\nabla^2M\;,\\
\frac{\delta\mathcal{F}_c}{\delta \tilde{G}_{ij}}&=&
\chi \tilde{G}_{ij}-K_G\nabla^2\tilde{G}_{ij}
\eea

It is useful to consider a simplified form of the equations obtained by retaining only lowest order terms in fields and gradients. In this case the Stokes equation and the equations for the shape fields can be written in the explicit form
\be
\Gamma\mathbf{v}=-\bm\nabla p+\eta\nabla^2 \mathbf{v}+ \eta_b\bm\nabla\bm\nabla\cdot\mathbf{v}+\bm\nabla\cdot\bm\sigma^G\;,
\ee
\be
\frac{d}{dt}M= 2\frac{R}{M}\tilde{G}_{ij} D_{ij} -\frac{1}{\gamma_M}\left[\alpha+\beta M^2\right]M +D\nabla^2 M\;,
%-\frac{\chi}{\gamma_2}\tilde{G}_{ij}\tilde{G}_{ij}\;,
\label{eq:Meq-final}
\ee
\bea
\frac{D}{Dt}\tilde{G}_{ij}&=&R D_{ij}+\tilde{G}_{ik}D_{kj}+D_{ik}\tilde{G}_{kj}-\delta_{ij}\tilde{G}_{kl} D_{kl}\notag\\
&-& r M \tilde{G}_{ij}
+D_G\nabla^2 \tilde{G}_{ij} \;,
\label{eq:Gij-final}
\eea
where
$D=\Gamma_M K$, $D_G=\Gamma K_G$, $r=\chi/\gamma_G+\alpha/\gamma_2$ and
\be
\sigma_{ij}^G=2R(\alpha+\beta M^2)\tilde{G}_{ij}+\frac12\delta_{ij}\chi S_G^2\;.
\ee
The single-cell anisotropy field $M$ here plays the role of tissue fluidity. The first term on the RHS of Eq.~\eqref{eq:Meq-final}  captures the fact that shear deformations, coupled to local cell alignment, can increase cell anisotropy, driving fluidification. The second term describes relaxation to the ground state controlled by the tissue free energy, with a cost for spatial variations in local fluidity controlled by the stiffness $D$.    The reactive terms in Eq.~\eqref{eq:Gij-final} describe flow alignment of elongated cell shape.  The term proportional to $r$ describes changes of cell shape tensor due to dissipative processes, such as topological rearrangements, at a rate   proportional to  the tissue fluidity $M$.  The last term in Eq.~\eqref{eq:Gij-final} describes the stiffness against deformations of local cell alignment.
%\ah{Note that reactive terms on the rhs Eq. 54 indicates that a homogeneous deformation of tissue induces cell elongation, weighted by cell perimeter density and anisotropy. This matches previous models where such a term is input by hand \cite{popovic2017}. }\par
%\ah{Additionally, Eq 54 tells us that the pure shear flow rate picks up contributions from the flow alignment via the coupling to $\tilde{G}_{ij}$ as well as being renormalized by energetic costs from shape change due to interactions.}

Finally,  in a confluent tissue the cell number density $n=\rho/m_c$ is slaved to the mean cell area $\langle A_\alpha\rangle$ with $n=1/\langle A_\alpha\rangle$. For cells that are only slightly deformed from regular polygons, $\langle A_\alpha\rangle\simeq\sqrt{\det[\mathbf{G}]}\approx \text{Tr}[\mathbf{G}]$, where we have used Eq.~\eqref{eq:id1}. The density equation, Eq~\eqref{eq:rhoeq-tissue}, can therefore equivalently be written as an equation for the cell area or for $|G|\equiv \det[\mathbf{G}]$, given by
\be
\left(\partial_t+\mathbf{v}\cdot\bm\nabla\right)|G|=|G|\bm\nabla\cdot\mathbf{v}\;.
\label{eq:detG}
\ee

\section{Conclusion}
Using the Poisson bracket formalism, we have derived  hydrodynamic equations for a fluid of deformable particles in two dimensions. Shape fluctuations  are described by two continuum fields: (i) a coarse-grained scalar field that captures single-particle anisotropy, and (ii) a shape tensor field that quantifies both particle  elongation and nematic alignment of elongated particles. 

We have specifically applied the model to sheets of dense biological tissue, where single-cell anisotropy was recently identified as the order parameter for a solid-liquid transition driven by the interplay of cortex contractility and cell-cell adhesion~\cite{bi2015density,bi2016motility}. In other words, in confluent tissue single-cell anisotropy is effectively an experimentally accessible measure of the rheological
properties of the tissue, with isotropic cell shapes
identifying the solid or jammed state and anisotropic shapes
corresponding to a liquid. Previous work has examined the dynamics of a coarse-grained cell  shape tensor and its coupling to mechanical stresses~\cite{ishihara2017cells}. This work did not, however, distinguish between a tissue of elongated, but isotropically oriented cells and one were the cells and elongated and also aligned in a state with nematic liquid crystalline order.
The new ingredient of our work is to distinguish the dynamics of tissue fluidity, as quantified by the single-cell anisotropy field, from that of cell alignment, and examine the interplay between flow, which can be either externally applied or induced by internal active processes, fluidity and nematic order of cell shapes.  Our equations hence provide a starting point for quantifying the rheology of biological tissue.  
Future extension needed to develop a complete framework of tissue rheology include the coupling to the dynamics of polarized cell motility and the inclusion of structural rearrangements arising from cell division and death.

Finally, the equations developed here provide a general hydrodynamic model for any fluid of deformable particles, capable of accounting for both the dynamics of shape deformations and density changes.

%\ah{In our work we have derived a continuum model for tissues where variable cell shape contributes to flow alignment and elastic stress. The strength of our works lies in that the PB bracket formalism stipulates the generic kinematic coupling between cell shape and hydrodynamic modes independent of a particular free energy. } 
\label{sec:end}

\acknowledgments
MCM thanks to Max Bi, James Cochran, Suzanne Fielding and Holger Stark for illuminating discussions. This work was supported by the National Science Foundation through award DMR-1938187.

\appendix
\section{Useful  identities}
\label{app:id}

The eigenvalues of a $2\times2$ symmetric matrix $G_{ij}$ are given by
\be
\lambda_{1,2}=\frac12(G_{xx}+G_{yy})\pm\frac12\sqrt{(G_{xx}-G_{yy})^2+4G_{xy}^2}\;,
\ee
and
\bea
&&\lambda_1-\lambda_2=\sqrt{(G_{xx}-G_{yy})^2+4G_{xy}^2}\;,\\
&&\lambda_1\lambda_2=G_{xx}G_{yy}-G_{xy}^2\;.
\eea
We can then show that the following identities apply
\bea
&&(\lambda_1-\lambda_2)^2=[\text{Tr}\mathbf{G}]^2-4\det\mathbf{G}\label{eq:id1}\;,\\
&&(\lambda_1-\lambda_2)^2=2\text{Tr}[\mathbf{G}^2]-[\text{Tr}\mathbf{G}]^2=2\text{Tr}[\mathbf{\tilde{G}}^2]\label{eq:id2}\;.
\eea

Finally, for a regular polygon, $G_{ij}$ is always diagonal and $\lambda_1=\lambda_2=\lambda$.
In this case Eq.~\eqref{eq:id1} gives $\text{Tr}\mathbf{G}=2\sqrt{\det\mathbf{G}}$.
For small deformations from a regular polygon  $\text{Tr}\mathbf{G}\sim2\sqrt{\det\mathbf{G}}$,
which implies that we can think of $\text{Tr}\mathbf{G}$ as either a measure of square of cell perimeter or a measure of cell area.
%%%%%%%%%%%%%%%%%%%%%%%%%%%%%%%%%%%%%%%%%%%%%%%%%%%%%
\section{Elastic Stress and Pressure}
\label{app:pressure}
It is convenient to rewrite some of the term in the reactive part $\mathbf{V}^g$ of the momentum density equation given in Eq. ~\eqref{eq:Vg} to express them as gradients of pressure and an elastic stress. The goal is to rewrite the following terms
%\onecolumngrid
\begin{equation}
\begin{aligned}
\delta V_i^g\equiv-\rho\partial_i\frac{\delta\mathcal{F}_V}{\delta\rho}
+(\partial_iM)\frac{\delta\mathcal{F}_V}{\delta M}+\left(\partial_iG_{kl}\right)\frac{\delta\mathcal{F}_V}{\delta G_{kl}}\;.
\label{eq:rewrite}
\end{aligned}
\end{equation}
By relating functional derivatives of $\mathcal{F}_V$ t to derivatives of the free energy density $f$, which is a function of the hydrodynamic fields and their gradients, we can write
%
%\onecolumngrid
\bea
&&-\rho\nabla_i \frac{\delta\mathcal{F}_V}{\delta\rho}
 = -\nabla_i\left(\rho\frac{\partial f}{\partial \rho}\right)+\frac{\partial f}{\partial \rho}\nabla_i\rho\;,
\eea
 %+\nabla_j\left(\nabla_i \frac{\partial f}{\partial \nabla_j \rho}}\right)+\nabla_j\left(\rho\nabla_i \frac{\partial f}{\partial \nabla_j \rho}}\right)\\
\bea
 &&(\nabla_i M)\frac{\delta\mathcal{F}_V}{\delta M}=(\nabla_i M)\left(\frac{\partial f}{\partial M}-\nabla_j\frac{\partial f}{\partial \nabla_j M}\right)= \frac{\partial f}{\partial M}\nabla_i M \notag\\
 &&
-\nabla_j\left[(\nabla_i M)\frac{\partial f}{\partial \nabla_j M}\right]
+\frac{\partial f}{\partial \nabla_j M}\nabla_i(\nabla_j M)\;,
\eea
\bea
&&\left(\nabla_iG_{kl}\right)\frac{\delta\mathcal{F}_V}{\delta G_{kl}} =\left(\nabla_iG_{kl}\right)\left(\frac{\partial f}{\partial G_{kl}}-\nabla_j\frac{\partial f}{\partial \nabla_jG_{kl}}\right)\notag\\
&&= \frac{\partial f}{\partial G_{kl}}\nabla_i G_{kl} 
-\nabla_j\left[(\nabla_i G_{kl})\frac{\partial f}{\partial \nabla_jG_{kl}}\right]\notag\\
&&+\frac{\partial f}{\partial \nabla_jG_{kl}}\nabla_i(\nabla_jG_{kl}) \;.
\label{eq:rewrite2}
\eea
%\twocolumngrid
%
Combining these three terms, and using
\bea
\nabla_if=\frac{\partial f}{\partial \rho}\nabla_i\rho+\frac{\partial f}{\partial M}\nabla_iM+\frac{\partial f}{\partial \nabla_jM}\nabla_i(\nabla_jM)\notag \\
+\frac{\partial f}{\partial G_{kl}}\nabla_i G_{kl}
+\frac{\partial f}{\partial \nabla_jG_{kl}}\nabla_i(\nabla_jG_{kl})
\eea
%\twocolumngrid
we can write
\be
\delta V_i^g=-\nabla_ip+\nabla_j\sigma_{ij}^E
\ee
in terms of the pressure $p$ and an elastic stress $\sigma_{ij}^E$, given by
\bea
p&=&\rho\frac{\partial f}{\partial\rho}-f
\label{eq:pressure}\;,\\
\sigma_{ij}^E&=&
%-\frac{\partial f}{\partial \nabla_i \rho} \nabla_j \rho 
- \frac{\partial f}{\partial \nabla_j M}\nabla_i M - \frac{\partial f}{\partial \nabla_j G_{kl}}\nabla_i G_{kl}\;.
\label{eq:sigmaE}
\eea
The stress $\sigma_{ij}^E$  plays the role of the Erickssen stress of nematic liquid crystals.

%%%%%%%%%%%%%%%%%%%%%%%%%%%%%%%%%%%%%%%%%%%%%%%%%%%%%%%%%%%%%%%%%%%%%%%%
\section{Evaluation of Poisson Brackets}
\label{app:PB}

First we show the details of the calculation of the fundamental PB
$\{G_{ij}^\alpha\delta(\br-\br^\alpha),g_k(\br')\}$.
To evaluate the PB  we use the following
\begin{align}
&\frac{\partial\Delta x_i^{\alpha\nu}}{\partial x_j^{\beta\mu}}=\delta^{\alpha\beta}\delta_{ij}\left(\delta^{\mu\nu}-\frac{1}{n}\right)\;,\\
&\frac{\partial }{\partial x_j^{\beta\mu}}\delta(\br-\br^\alpha)=-\frac{\delta^{\alpha\beta}}{n}\partial_j\delta(\br-\br^\alpha)\;,\\
&\delta(\br-\br^\alpha)=\delta(\br-\br^{\alpha\mu}-\Delta\br^{\alpha\mu})\notag\\
&\;\;\;\;\;\;~~~~~~~=\delta(\br-\br^{\alpha})-\Delta x_i^{\alpha\mu}\partial_i\delta(\br-\br^{\alpha\mu})+\mathcal{O}(\Delta x^2\nabla^2)\;.
\end{align}
We  write
\be
\{G_{ij}^\alpha\delta(\br-\br^\alpha),g_k(\br')\}=
-\sum_{\beta,\nu}\frac{\partial G_{ij}^\alpha\delta(\br-\br^\alpha)}{\partial x^{\beta\nu}_k}\delta(\br'-\br^{\beta\nu})\;.
\label{eq:AC1}
\ee
Then
\bea
\frac{\partial G_{ij}^\alpha\delta(\br-\br^\alpha)}{\partial x^{\beta\nu}_k}&&=
-\frac{1}{n}\delta^{\alpha\beta}G_{ij}^\alpha\partial_k\delta(\br-\br^\alpha)\notag\\
&&+\frac{1}{n}\delta^{\alpha\beta}\delta(\br-\br^\alpha)
\left(\delta_{ik}\Delta x^{\alpha\nu}_j+\delta_{jk}\Delta x^{\alpha\nu}_i\right)\;.
\label{eq:AC2}
\eea
Inserting Eq.(C5) %~\eqref{eq:AC2}
into Eq.(C4) %~\eqref{eq:AC1}
and using that $\sum_\mu\Delta \br^{\alpha\mu}=0$, we obtain
%Using \eqref resulted in big gaps between words, fixed it by hand.
\bea
&&\{G_{ij}^\alpha\delta(\br-\br^\alpha),g_k(\br')\}=
G_{ij}^\alpha\left[\partial_k\delta(\br-\br^\alpha)\right]\frac{1}{n}\sum_\nu \delta(\br'-\br^{\alpha\nu})\notag\\
&&-\delta(\br-\br^\alpha)\frac{1}{n}\sum_\nu\left[\delta_{ik}\Delta^{\alpha\nu}_j+\delta_{jk}\Delta^{\alpha\nu}_i\right]\delta(\br'-\br^{\alpha\nu})\;.
\eea
Finally, using
\bea
\delta(\br-\br^{\alpha\nu})&&=\delta(\br-\br^\alpha-\Delta\br^{\alpha\nu})\notag\\
&&\approx \delta(\br-\br^\alpha)-\Delta x^{\alpha\nu}_k\partial_k\delta(\br-\br^\alpha)\;,
\eea
we obtain
\bea
\{G_{ij}^\alpha\delta(\br-\br^\alpha),g_k(\br')\}=
G_{ij}^\alpha\delta(\br'-\br^\alpha)\partial_k\delta(\br-\br^\alpha)\notag\\
+\delta(\br-\br^\alpha)\left[\delta_{ik}G_{jl}^\alpha+\delta_{jk}G_{il}^\alpha\right]
\partial'_l\delta(\br'-\br^\alpha)\;.
\label{eq:F1-PB}
\eea
From this one can immediately obtain Eq.~\eqref{eq:PB-Gij}.
%\twocolumngrid

To evaluate the PB $\{M(\br),g_k(\br')\}$ we let 
$G_{ij}^\alpha=\frac{I_\alpha}{2}\delta_{ij}+\tilde{G}^\alpha_{ij}$, with $\tilde{G}^\alpha_{ij}=\Delta_\alpha\left(\nu^\alpha_i\nu^\alpha_j-\frac12\delta_{ij}\right)$ and use the following identities
\bea
\tilde{G}_{ik}^\alpha \tilde{G}_{kj}^\alpha &=&\frac{\Delta_\alpha^2}{4}\delta_{ij}\;,\\
\tilde{G}_{ik}^\alpha \tilde{G}_{ki}^\alpha &=&\frac{\Delta_\alpha^2}{2}\;.
\eea
We can then write
\be
\Delta_\alpha\{\Delta_\alpha\delta(\br-\br^\alpha),g_k(\br')\}
=2\tilde{G}^\alpha_{ij}\{\tilde{G}^\alpha_{ij}\delta(\br-\br^\alpha),g_k(\br')\}\;.
\ee
Using Eq.~\eqref{eq:F1-PB}, we find
%\onecolumngrid

\bea
&&\{\tilde{G}_{ij}^\alpha \delta(\br-\br^\alpha),g_k(\br')\}=
\tilde{G}_{ij}^\alpha\delta(\br'-\br^\alpha)\partial_k\delta(\br-\br^\alpha)\notag\\
&&+\delta(\br-\br^\alpha)\left[\delta_{ik}\tilde{G}_{jl}^\alpha+\delta_{jk}\tilde{G}_{il}^\alpha-\delta_{ij}\tilde{G}_{kl}^\alpha\right]
\partial'_l\delta(\br'-\br^\alpha)\notag\\
&&+\frac{I_{\alpha}}{2}\left(\delta_{ik}\delta_{jl}+\delta_{jk}\delta_{il}-
%\eea
\delta_{ij}\delta_{kl}\right)\delta(\br-\br^\alpha)\partial'_l\delta(\br'-\br^\alpha). 
%\eea
\label{eq:F-PB}
\eea
and
\bea
\{\Delta_\alpha\delta(\br-\br^\alpha)&&,g_k(\br')\}=
\Delta_\alpha\delta(\br'-\br^\alpha)\partial_k\delta(\br-\br^\alpha)\notag\\
&&+\Delta_\alpha\delta(\br-\br^\alpha)\partial'_k\delta(\br'-\br^\alpha)\notag\\
&&+\frac{2I_\alpha\tilde{G}_{kl}^\alpha}{\Delta_\alpha}\delta(\br-\br^\alpha)\partial'_l\delta(\br'-\br^\alpha)\;.
\eea

\twocolumngrid

The PB $\{M(\br),g_k(\br')\}$ is then given by

\bea
\{M(\br),&&g_k(\br')\}=\delta(\br-\br')\partial_kM(\br)\notag\\
&&-2\left[\sum_\alpha\frac{I_\alpha\tilde{G}_{kl}^\alpha}{\Delta_\alpha}\delta(\br-\br^\alpha)\right]\partial_l\delta(\br-\br')
\eea
and involves a new field
\be
\sum_\alpha \frac{I_\alpha\tilde{G}_{kl}^\alpha}{\Delta_\alpha}\delta(\br-\br^\alpha)=\sum_\alpha I_\alpha\left(\nu^\alpha_i\nu^\alpha_j-\frac12\delta_{ij}\right)\delta(\br-\br^\alpha)\;.
\ee
We will need to make approximations to close the equations.
We will approximate as follows
\be
\sum_\alpha \frac{I_\alpha\tilde{G}_{kl}^\alpha}{\Delta_\alpha}\delta(\br-\br^\alpha)\approx  \frac{R(\br)\tilde{G}_{ij}(\br)}{M(\br)}\;.
\ee

%Finally, it is also of interest to consider  the PB of the normalized  shape anisotropy,
%\be 
%\hat{m}(\br,t)= \sum_{\alpha} %\frac{\lambda_1^a-\lambda_2^a}{\lambda_1^a+\lambda_2^a} %\delta(\br-\br_{\alpha}(t))\;.
%\ee
%This can be evaluated using the product rule
%
%\bea
%\{\frac{\Delta_{\alpha}}{I_{\alpha}}\delta(\br-\br^{\alpha})%, &&g_{k}(\br')\}= \frac{1}{I_{\alpha}} %\{\Delta_{\alpha}\delta(\br-\br^{\alpha}), g_{k}(\br')\} %\notag\\
%&&- \frac{\Delta_{\alpha}}{I_{\alpha}^2}\{I_{\alpha}\delta(\%br-\br^{\alpha}), g_{k}(\br')\}\;.
%\eea
%
%\mcm{The following is at best a part of the equation, and does not look correct. Also, We did not define $Q_{ij}$ could you please write the correct equation here and write in terms of quantities that are defined?}
%The continuum result is 
%
%\be
%\frac{d M}{dt} = (1-M^2)Q_{ij}(\nabla_i v_j+ \nabla_j %v_i)\;.
%\ee
%
%%%%%%%%%%%%%%%%%%%%%%%%%%%%%%%%%%%%%%%%%%%%%%%%%%%%%%%%%%%%

\section{Mean Field theory of Vertex Model}
\label{app:MFT}

Following \cite{czajkowski2018hydrodynamics}, we construct a mean-field free energy  by rewriting the \emph{single-cell} Vertex model energy in terms of the cell anisotropy parameter $M_\alpha$. 
Let us define
\bea
&&M_\alpha=\lambda_1^\alpha-\lambda_2^\alpha\;,\\
&&R_\alpha=\lambda_1^\alpha+\lambda_2^\alpha\;,
\eea
which gives
$\lambda_{1,2}^\alpha=(R_\alpha \pm M_\alpha)/2$.
Equations \eqref{eq:An} and \eqref{eq:Pn} are exact for regular polygons, but also hold approximately true for slightly deformed polygons where the shape tensor remains diagonal and  $M_\alpha/R_\alpha\ll1$.
We can then write
%\mcm{are these equations correct?}
%
\bea
P_\alpha&\approx& 2n \sin\left(\frac{\pi}{n}\right)\left(\sqrt{\lambda_1^\alpha+\lambda_2^\alpha}\right)\equiv\nu(n)\sqrt{R_\alpha}\;,\\
A_\alpha&=&n \sin\left(\frac{2\pi}{n}\right)\sqrt{\lambda_1^\alpha\lambda_2^\alpha}\equiv \frac{\mu(n)}{2}\sqrt{R_\alpha^2-M_\alpha^2}\;.
%M&&= \lambda_1-\lambda_2
\eea

The single-cell energy can then be written in terms of $R_\alpha$ and $M_\alpha$ as
\bea
E_\alpha=\frac{K_A}{2}\left(\frac{\mu}{2}\sqrt{ R_\alpha^2-M_\alpha^2}-A_0\right)^2  
+\frac{K_P}{2}(P_\alpha-P_0)^2\;.
\eea
Expanding for $M_\alpha/R_\alpha\ll1$, we can write 
\bea
E_\alpha=&&
\frac{K_A}{2}\left[\left(\frac{\mu A_0}{2R_\alpha}-\frac{\mu}{2}\right)M_\alpha^2
+\frac{\mu A_0}{8R_\alpha^3}M_\alpha^4
+\left(\frac{\mu R_\alpha}{2}-A_0\right)^2\right]\notag\\
&&+\frac{K_P}{2}\left(\nu\sqrt{R_\alpha}-P_0
\right)^2\;.
\label{eq:Ea-mf}
\eea
We now assume that the cell perimeter is constant, or $P_\alpha =P_0$, hence $R_\alpha=P_0^2 / \nu^2$.
Substituting into Eq.~\eqref{eq:Ea-mf}, we can rewrite the single-cell energy density $e_\alpha=E_\alpha/A_0$ as
\bea
e_\alpha
=  e_0+\frac{1}{2}\alpha (n,p_0)\left(\frac{M_\alpha}{{A_0}}\right)^2 + \frac{1}{4}\beta(n,p_0)\left(\frac{M_\alpha}{{A_0}}\right)^4 \;,
\label{eq:e-fin}
\eea
where $e_0$ is a constant and 
\bea
\alpha(n,p_0)&=& \frac{\kappa_A A_0^2 \mu^2}{4p_0^2} \left({p_0^{*}}^2-p_0^2\right)
%\mcm{\frac{n^2 \sin\left(\frac{2\pi}{n}\right)^2}{18 A_0^2 p_0^2}\left(4n\tan \left(\frac{\pi}{n}\right)-p_0^2\right)=...(p_0^*-p_0)}\;,
\label{eq:alpha}\;,\\
\beta(n,p_0)&=& \frac{\kappa_a A_0^4\mu\nu^6}{4p_0^6} 
%\mcm{\frac{n^6 \sin ^6\left(\frac{\pi }{n}\right) \cos ^2\left(\frac{\pi }{n}\right)}{4
  % A_0^4 p_0^4}}
   \;,
   \label{eq:beta}
\eea
with $p_0=P_0/\sqrt{A_0}$  the shape index, our tuning parameter. Also, $\alpha(n,p_0)$ has been written in terms of the critical shape index, 
\bea
p_0^*=\nu\sqrt{\frac{2}{\mu}}
%=\frac{2\sqrt{2} n\sin(\pi/n)}{\sqrt{n\sin(2\pi/n)}}
= \sqrt{4n\tan(\pi/n)} \;.
\label{eq:p0star}
\eea
%given by $p_0^*= \sqrt{4n\tan \left(\frac{\pi}{n}\right)}$. 
%Note that  the anisotropy field of an individual cell $M_\alpha$ has dimensions of square of length. 
The value of critical target shape parameter $p_0^*$ depends on the specific undeformed polygonal shape, with $p_0^*=4$ for squares and $p_0^*=2\sqrt{2\sqrt{3}}\approx 3.722$ for hexagons. Eq.~\eqref{eq:alpha} shows explicitly that $\alpha$ changes sign at $p_0=p_0^*$, while $\beta>0$. For $\alpha>0$ the stable ground state has $M=0$ and corresponds to a solid-like state of isotropic cells. For $\alpha<0$ the stable ground state is a fluid of anisotropic cells, with $M/A_0=\pm\sqrt{-\alpha/\beta}$. At $\alpha=0$ the system undergoes spontaneous symmetry breaking and fluidizes, choosing one of two equivalent axial direction along which to elongate. Here we have defined  $M $ as positive by assuming $\lambda_1>\lambda_2$, hence breaking from the outside the Ising symmetry of the model.
Finally, it was shown in Ref.~\cite{czajkowski2018hydrodynamics} that the quartic form given in Eq.~\eqref{eq:e-fin} is also obtained by assuming constant cell area, albeit with different expressions for the coefficients $\alpha$ and $\beta$. In both cases the coefficient $\alpha$ changes sign at $p_0=p_0^*$ and the behavior near the transition is unaffected by the approximation used. 
%%%%%%%%%%%%%%%%%%%%%%%%%%%%%%%%%%%%%%%%%%%%%%%%%%%%%%

%%%%%%%%%%%%%%%%%%%%%%%%%%%%%%%%%%%%%%%%%%%%%%%%%%%%%%%%%%%%
\bibliographystyle{unsrt}
\bibliography{PBref}

\end{document}